\documentclass{revtex4}

\usepackage{amsfonts}

\begin{document}
\draft

\title{Harmonic analysis on a Galois field and its subfields}

\author{A. Vourdas\\
Department of Computing,\\
University of Bradford, \\
Bradford BD7 1DP, United Kingdom}

\begin{abstract}
Complex functions $\chi (m)$ where $m$ belongs to a Galois field $GF(p^ \ell)$, are considered.
Fourier transforms, displacements in the  $GF(p^ \ell) \times GF(p^ \ell)$ phase space and symplectic 
$Sp(2,GF(p^ \ell))$ transforms
of these functions are studied. It is shown that the formalism inherits many features from the theory of Galois fields.
For example, Frobenius transformations are defined which leave fixed all functions $h(n)$ where $n$ belongs to a subfield 
$GF(p^ d)$ of the $GF(p^ \ell)$. The relationship between harmonic analysis (or quantum mechanics) on $GF(p^ \ell)$
and  harmonic analysis on its subfields, is studied.
\end{abstract}

\maketitle
{\bf key words: phase space methods, Heisenberg-Weyl group, Galois fields\\
MSC: 81S30, 42C30,13B05, 12F10}

\section{Introduction}

Quantum mechanics with complex wavefunctions $\chi (m)$ where $m$ belomgs to the ring ${\mathbb Z}_d$ (the integers modulo $d$) has been studied 
by Weyl \cite{1} and Schwinger \cite {2-1,2-2}
and later by many authors (a review with the relevant literature has been presented in \cite{3}).
Fourier transforms and displacements in phase space (which is the toroidal lattice ${\mathbb Z}_d \times {\mathbb Z}_d$)
have been studied extensively.

When $d$ is equal to a prime number number $p$, we get stronger results.
This is due to the fact that the ${\mathbb Z}_p$ is a field and the corresponding phase space
${\mathbb Z}_p \times {\mathbb Z}_p$ is a finite geometry \cite{4-1,4-2,4-3}. In this case there is nothing special with the
`position-momentum' (or `time-frequency') directions in phase space;
but all results can be proved with respect to the other directions in the finite geometry.
Symplectic transformations are well defined and form the $Sp (2,{\mathbb Z}_p)$ group.

In recent work \cite{5-1,5-2,5-3} we have studied in a quantum mechanical context
how we can go from ${\mathbb Z}_p$ to a bigger Galois field $GF(p^ \ell)$.
Related work has also been reported in \cite{6,7,8,8a}.

With a very different motivation there has been a lot of related work in the context of  
mutually unbiased bases in quantum systems with $d$-dimensional Hilbert space.
It is known that the number of such bases is less or equal to $d+1$; and that when $d$ is the power of a prime, it is equal to $d+1$.
This problem has recently been studied extensively in the quantum mechanical literature \cite{m1-1,m1-2,m1-3,m2,m3-1,m3-2,
m4-1,m4-2,m4-3,m5-1,m5-2,m6-1,m6-2,m7-1,m7-2}.

In this paper we continue the work of \cite{5-1,5-2,5-3} with emphasis on the mathematical aspects. We use ideas from field extension to go 
from harmonic analysis (or quantum mechanics) on ${\mathbb Z}_p$, to harmonic analysis on $GF(p^ \ell)$.
This mathematical structure inherits many features from the theory of field extension.
Frobenius transformations and Galois groups in the context of finite fields, have counterparts in the present context.
For example, we define Frobenius transforms which map the values of a function at Galois conjugates, 
($\chi (m)$, $\chi (m^p)$, $\chi (m^{p^2})$, etc) to each other.
We also discuss Galois groups which in the present  context are transformations in the Hilbert space $H$ 
of functions $\chi (m)$, which leave fixed all functions in the subspace ${\mathfrak H}_d$ comprised of functions
$h(n)$ where $n$ belongs to a subfield  $GF(p^ d)$ of $GF(p^ \ell)$.

In section II we consider complex functions on a Galois field and we discuss the basic formalism.
In section III we present Fourier transforms in both the space $H$ and its subspace ${\mathfrak H}_d$.
In section IV we discuss Frobenius transforms and the corresponding Galois groups, in the present context.
In section V we study displacements and the Heisenberg-Weyl group, in both the space $H$ and its subspace ${\mathfrak H}_d$. 
In section VI we consider symplectic transformations and the $Sp(2,GF(p^\ell))$ group.
We conclude in section VII with a discussion of our results.

\subsection{Notation}

The elements of the Galois field $GF(p^\ell)$ can be written as polynomials
\begin{eqnarray}\label{098}
m=m_0+m _1\epsilon +...+m _{\ell-1}\epsilon ^{\ell-1}
;\;\;\;\;\;\;m _0, m _1,...,m _{\ell-1}\in {\mathbb Z}_p 
\end{eqnarray}
They are defined modulo an irreducible polynomial of degree $\ell$:
\begin{eqnarray}\label{cha}
P(\epsilon)\equiv c _0+ c _1\epsilon +...+c _{\ell-1}\epsilon ^{\ell-1}+\epsilon ^\ell
;\;\;\;\;\;\;c _0, c _1,...,c _{\ell-1}\in {\mathbb Z}_p 
\end{eqnarray}
Different irreducible polynomials of the same degree $\ell$ lead to isomorphic finite fields. 
We refer to the $m _0, m _1,...,m _{\ell-1}$ as the Galois components of $m$
in the basis $1,\epsilon,...,\epsilon ^{\ell -1}$.

The $m$, $m ^p$,...,$m ^{p^{\ell-1}}$ are Galois conjugates.
The Frobenius map is
\begin{eqnarray}\label{74}
\sigma (m)=m^p;\;\;\;\;\;\;\sigma ^\ell=1
\end{eqnarray}
The 
\begin{eqnarray}\label{gal19}
{\rm Gal}[GF(p^\ell)/{\mathbb Z}_p]=\{{\bf 1},\sigma ,..., \sigma ^{\ell -1}\}  
\end{eqnarray}
is a cyclic group of order $\ell$. It comprises of all automorphisms of $GF(p^\ell)$
which leave the elements of the subfield ${\mathbb Z}_p$ fixed.

If $d$ is a divisor of $\ell$ (which we denote as $d|\ell$) the $GF(p^d)$ is a subfield of $GF(p^ \ell)$.
The 
\begin{eqnarray}\label{gal1}
{\rm Gal}[GF(p^\ell)/GF(p^d)]=\{{\bf 1},\sigma ^d,..., \sigma ^{\ell -d}\}  
\end{eqnarray}
is a cyclic group of order $\ell/d$ and is a subgroup of ${\rm Gal}[GF(p^\ell)/{\mathbb Z}_p]$. 
It comprises of all automorphisms of $GF(p^\ell)$
which leave the elements of the subfield $GF(p^d)$ fixed.

The trace of $m\in GF(p^\ell)$ is defined as:
\begin{eqnarray}\label{777}
{\rm Tr}(m)=m +m^p+...+m ^{p^{\ell-1}};\;\;\;\;\;\;\;{\rm Tr}(m)\in {\mathbb Z}_p;\;\;\;\;\;\;m\in GF(p^\ell)
\end{eqnarray}
When $m$ belongs to the subfield $GF(p^d)$ we make the distinction 
between the trace with regard to the extension from ${\mathbb Z}_p$
to $GF(p^\ell)$ given in Eq.(\ref{777}) and the trace with regard to the extension from ${\mathbb Z}_p$
to $GF(p^d)$ given by
\begin{eqnarray}\label{776}
{\mathfrak Tr}(m)=m +m^p+...+m ^{p^{d-1}};\;\;\;\;\;\;\;{\mathfrak Tr}(m)\in {\mathbb Z}_p;\;\;\;\;\;\;m\in GF(p^d)
\end{eqnarray}
Indices in the notation of the trace which indicate the extension, are sometimes used in the literature.
For simplicity we use ${\mathfrak Tr}$ in Eq.(\ref{776}) and ${\rm Tr}$ in Eq.(\ref{777}).
It is easily seen that
\begin{eqnarray}\label{775}
{\rm Tr}(m)=\frac{\ell}{d}\;{\mathfrak Tr}(m);\;\;\;\;\;\;m\in GF(p^d)
\end{eqnarray}
In the special case that $\ell/d$ is a multiple of the prime $p$, all $m$ in $GF(p^d)$ have ${\rm Tr}(m)=0$.
We note that this does not contradict a theorem which states that in every finite field there exists at least one element
with non-zero trace. According to this theorem it is impossible to have ${\mathfrak Tr}(m)=0$
for all $m$ in $GF(p^d)$. But it is possible to have  ${\rm Tr}(m)=0$ 
for all $m$ in $GF(p^d)$ because this is a different trace.

Following \cite{5-1} we introduce the $\ell\times\ell$ matrices
\begin{eqnarray}
g_{\lambda \kappa}\equiv{\rm Tr}\left (\epsilon ^{\lambda +\kappa}\right )
;\;\;\;\;\;G\equiv g^{-1};\;\;\;\;\;g_{\lambda \kappa},G_{\lambda \kappa}\in {\mathbb Z}_p
\end{eqnarray}
where $\lambda$ and $\kappa$ take values from $0$ to $\ell -1$.
Using them we define the dual basis  $E_0, E_1,...,E_{\ell -1}$, as:
\begin{eqnarray}
E_{\kappa}=\sum _{\lambda } G_{\kappa \lambda}\epsilon ^{\lambda };\;\;\;\;\;\;\;\;
{\rm Tr}(\epsilon ^\kappa E_{\lambda})=\delta _{\kappa \lambda}
\end{eqnarray}
A number $m \in GF(p^\ell)$ can be expressed in the two bases as:
\begin{eqnarray}\label{kkk}
m &=&\sum _{\lambda=0}^{\ell-1}m _ {\lambda} \epsilon ^{\lambda}=
\sum _{\lambda=0}^{\ell-1}{\overline m} _\lambda E_{\lambda }\nonumber\\
m _\lambda &=&{\rm Tr}[m E_{\lambda }];\;\;\;\;\;
{\overline m} _{\lambda} ={\rm Tr}[m {\epsilon} ^\lambda ]\nonumber\\
m _{\lambda }&=&\sum _{\kappa} G_{ \lambda \kappa}{\overline m} _{\kappa};\;\;\;\;\;\;
{\overline m} _{\lambda }=\sum _{\kappa} g_{ \lambda \kappa}m _{\kappa}
\end{eqnarray}
We refer to ${\bar m} _{\lambda }$ as the dual Galois components of $m$.
The trace of the product $\alpha \beta $ is given in terms of the components of these numbers as
\begin{eqnarray}\label{102}
{\rm Tr}(mn)&=&\sum _{\lambda,\kappa}g_{\lambda \kappa}m_\lambda n _\kappa =
\sum _{\lambda,\kappa}G_{\lambda \kappa}{\overline m} _\lambda {\overline n} _\kappa\nonumber\\&=&
\sum _{\lambda} m _\lambda {\overline n}_\lambda=\sum _{\lambda} {\overline m}_\lambda n _\lambda
\end{eqnarray}

\section{Complex functions on a Galois field and its subfields}

We consider the $p^\ell$-dimensional Hilbert space $H$ of complex vectors
$\chi(m)$ where $m\in GF(p^ \ell)$. The scalar product $(\chi, h)$ is defined as
\begin{eqnarray}	
(\chi, h)=\sum _{m}[\chi (m)]^*h(m)
\end{eqnarray} 
where the summation is over all $m\in GF(p^ \ell)$.
We define the projection operators ${\cal Q}_k$
\begin{eqnarray}\label{900}	
&&{\cal Q}_k(m,n)=1;\;\;\;{\rm if}\;\;m=n=k\nonumber\\
&&{\cal Q}_k(m,n)=0;\;\;\;{\rm otherwise}
\end{eqnarray}
They obey the relations
\begin{eqnarray}	
{\cal Q}_k {\cal Q}_m=\delta (k,m){\cal Q}_k;\;\;\;\;\;\;\;\sum _{k}{\cal Q}_k={\bf 1}
\end{eqnarray}
where $\delta $ is the Kronecker delta.

We have explained that when $d$ is a divisor of $\ell$  the $GF(p^d)$ is a subfield of $GF(p^ \ell)$.
The complex vectors $\chi (m)$ where $m\in GF(p^ d)$ belong to a subspace of $H$ which we call ${\mathfrak H}_d$.
For example, the subspace ${\mathfrak H}_1$ contains complex functions $f(m)$ where $m\in {\mathbb Z}_p$; and  
the subspace ${\mathfrak H}_ \ell=H$.
The projection operator $\Pi _d(m,n)$ from $H$ to ${\mathfrak H}_d$ is 
\begin{eqnarray}	
\Pi _d=\sum _{k\in GF(p^d)}{\cal Q}_k
\end{eqnarray} 
If $c$ is a divisor of $d$ then
\begin{eqnarray}	
c|d\;\rightarrow\;\Pi_c\Pi_d=\Pi_c
\end{eqnarray} 
The projection of the vector $\chi (m)$ to the space ${\mathfrak H}_d$ is
$\sum _n\Pi _d(m,n)\chi (n)$. 
We note that the subspaces ${\mathfrak H}_d$ depend on the basis that we choose.
With a unitary transformation $U$ we can go to a different basis and then 
we get different subspaces which we denote as $U{\mathfrak H}_d$ (the corresponding
projection operator is $U\Pi _d U^{\dagger}$).

An orthonormal basis in the Hilbert space $H$ is the functions
\begin{eqnarray}\label{ppp}
\phi _n(m)=(p^\ell)^{-1/2}\omega [-{\rm Tr}(nm)]
\end{eqnarray}
where $n\in GF(p^ \ell)$ and
\begin{eqnarray}\label{8fr}
\omega= \exp \left (i\frac{2\pi}{p}\right );\;\;\;\;\;\;\;\omega (\alpha )\equiv \omega ^{\alpha};\;\;\;\;\;\;\;\;\alpha \in {\mathbb Z}_p
\end{eqnarray}
Indeed we can easily show that for $k,n,m,r \in GF(p^\ell)$:
\begin{eqnarray}\label{23}
(\phi _n, \phi _r)=\delta (n,r)\nonumber\\	
\sum _n [\phi _n (m)]^* \phi _n (k)=\delta (m,k)
\end{eqnarray}

The properties of the trace lead to the relation
\begin{eqnarray}\label{ppp}
\phi _n(m)=\phi _{n^p}(m^p)=...=\phi _{n^{p^ {\ell -1}}}(m^{p^ {\ell -1}})
\end{eqnarray}

The Hilbert space $H$ can be written as the tensor product ${\cal H}\otimes ...\otimes {\cal H}$
where ${\cal H}$ are $p$-dimensional `component Hilbert spaces' of complex functions $g(m_i)$ where $m_i\in {\mathbb Z}_p$.
Indeed, the general function $\chi (m)$ can be written in terms of the Galois components of $m$ as $\chi (m_0,...,m_{\ell -1})$.
As an example of this, we use Eq.(\ref{102}) and show that
\begin{eqnarray}
\phi _n(m)&=&{\varphi} _{n_0}({\overline m}_0)...\varphi _{n_{\ell -1}}({\overline m}_{\ell -1})\nonumber\\&=&
{\varphi} _{{\overline n}_0}(m_0)...\varphi _{{\overline n}_{\ell -1}}(m_{\ell -1})
\end{eqnarray}
where
\begin{eqnarray}
{\varphi} _{\alpha }(\beta)=p^{-1/2}\omega (-\alpha \beta);\;\;\;\;\;\alpha ,\beta \in {\mathbb Z}_p
\end{eqnarray}
belongs to the Hilbert space ${\cal H}$.

\section{Fourier transform}

The Fourier matrix $F$ is the $p^\ell \times p^\ell$ matrix:
\begin{eqnarray}\label{410}
F(n,m)=(p^\ell)^{-1/2}\omega [{\rm Tr}(nm)];\;\;\;\;\;\;\;F^4={\bf 1};\;\;\;\;\;\;FF^{\dagger}={\bf 1}
\end{eqnarray}
The Fourier transform of a function $\chi(m)$ in $H$, is given by
\begin{eqnarray}\label{ppp}
{\tilde \chi }(n)=(\phi _n ,\chi)=\sum _m F(n,m)\chi (m)
\end{eqnarray}

Using Eq.(\ref{102}) we show that
\begin{eqnarray}\label{ppp1}
F(n,m)&=&{\cal F}({\overline n}_0,m_0)...{\cal F}({\overline n}_{\ell -1},m_{\ell -1})\nonumber\\
&=&{\cal F}(n_0,{\overline m}_0)...{\cal F}(n_{\ell -1},{\overline m}_{\ell -1})
\end{eqnarray}
where 
\begin{eqnarray}\label{pol}
{\cal F}(\alpha ,\beta)=p^{-1/2}\omega (\alpha \beta );\;\;\;\;\;\;\alpha ,\beta \in {\mathbb Z}_p
\end{eqnarray}
are Fourier transforms in the `component spaces' ${\cal H}$.
It is seen that the dual components of $m$ and the components of $n$ appear in 
Eq.(\ref{ppp1}) (or vice-versa). Therefore in general
\begin{eqnarray}
F(n,m)&\ne &{\cal F}(n_0,m_0)...{\cal F}(n_{\ell -1},m_{\ell -1})
\end{eqnarray}
This equation shows one of the differences between harmonic analysis on $GF(p^\ell)$ and harmonic analysis on
${\mathbb Z}_p\times ...\times {\mathbb Z}_p$. The former uses the Fourier transform $F(n,m)$ and the latter the
Fourier transform ${\cal F}(n_0,m_0)...{\cal F}(n_{\ell -1},m_{\ell -1})$.

The properties of the trace lead to the relation
\begin{eqnarray}\label{65}
F(n,m)=F(n^p,m^p)=...=F(n^{p^{\ell -1}},m^{p^{\ell -1}})
\end{eqnarray}

\subsection{Fourier transforms in ${\mathfrak H}_d$}

We consider the space ${\mathfrak H}_d$ (where $d$ is a divisor of $\ell$) and in analogy with Eq.(\ref{410}), we introduce the   
$p^d\times p^d$ Fourier matrix
\begin{eqnarray}\label{410a}
&&{\mathfrak F}(n,m)=(p^d)^{-1/2}\omega [{\mathfrak Tr}(nm)];\;\;\;\;\;\;n,m\in GF(p^d)\nonumber\\
&&{\mathfrak F}^4=\Pi _d;\;\;\;\;\;\;{\mathfrak F}{\mathfrak F}^{\dagger}=\Pi _d
\end{eqnarray}
We note that the trace of Eq.(\ref{776}) is used here, in contrast to Eq.(\ref{410}) where the trace of Eq.(\ref{777}) is used.
We compare  ${\mathfrak F}(n,m)$ with the matrix $\Pi _d F\Pi_d$. 
The  $\Pi _d F\Pi_d$ is a $p^\ell \times p^\ell$
matrix but only  $p^d\times p^d$ elements are non-zero.
Taking into account Eq.(\ref{775}) we show that for $n,m\in GF(p^d)$
\begin{eqnarray}\label{411}
(\Pi _d F\Pi_d)(n,m)=[{\mathfrak F}(n,m)]^{\ell/d}
\end{eqnarray}
This shows that the elements of $F$ with indices in $GF(p^d)$ are powers of the corresponding elements of ${\mathfrak F}$.
Of course, the matrix $F$ has other elements also, with indices in the set $GF(p^\ell)-GF(p^d)$.
In the special case that $\ell/d$ is a multiple of the prime $p$, Eq.(\ref{411}) shows that the $p^d\times p^d$ submatrix of $F(n,m)$
with indices in $GF(p^d)$, has all its elements equal to $p^{-\ell/2}$.

\subsection{Spectrum of Fourier transforms}
From  Eq.(\ref{410}) it follows that the eigenvalues of $F$ are $1,i,-1,-i$.
We express $F$ in terms of its eigenvalues and eigenvectors as:
\begin{eqnarray}\label{37}
&&F=\pi _0+i\pi _1-\pi _2-i\pi _3\nonumber\\
&&\pi _r \pi _s=\pi _{r }\delta (r,s);\;\;\;\;\;\;\;\pi _0+\pi _1+\pi _2+\pi _3={\bf 1};\;\;\;\;\;
r,s=0,1,2,3
\end{eqnarray}
Here $\pi _{\lambda }$ are orthogonal projectors to the eigenspaces corresponding to the various eigenvalues of $F$.
From Eq.(\ref{410}) it follows that:
\begin{eqnarray}\label{710}
\pi _r=\frac{1}{4}\left [{\bf 1}+(i^{-r}F)+(i ^{-r}F)^2+(i ^{-r}F)^3\right ];\;\;\;\;\;\;r =0,1,2,3
\end{eqnarray}

\section{Frobenius transform}

The Frobenius map of Eq.(\ref{74}) leads in the present context to the Frobenius transform.
We define the Frobenius matrix as 
\begin{eqnarray}\label{307}
{\cal G}(n,m)=\delta (n,m^p);\;\;\;\;\;{\cal G}^\ell ={\bf 1};\;\;\;\;\;\;{\cal G}{\cal G}^{\dagger}={\bf 1}
\end{eqnarray} 
The Frobenius transform of a function $\chi(m)$ in $H$, is given by
\begin{eqnarray}\label{307a}
({\cal G}\chi) (n)=\sum _m {\cal G}(n,m)\chi (m)=\chi (n^{p^{\ell -1}})
\end{eqnarray}
We note that ${\cal G}$ depends on the basis that we choose.
With a unitary transformation $U$ we can go to a different basis and then the Frobenius transform
becomes $U{\cal G}U^{\dagger}$.
The operator ${\cal G}$ commutes with the projection operators $\Pi _d$ (where $d$ is a divisor of $\ell $):
\begin{eqnarray}\label{30089}
[{\cal G},\Pi _d]=0
\end{eqnarray}

A direct consequence of our comments  about the group ${\rm Gal}[GF(p^\ell)/{\mathbb Z}_p]$ in Eq.(\ref{gal19}), is that the
\begin{eqnarray}\label{tt5}
{\rm Gal}(H/{\mathfrak H}_1)=\{{\bf 1}, {\cal G},...,{\cal G}^{\ell -1}\} 
\end{eqnarray}
is a cyclic group of order $\ell$.
These transformations leave all the functions in ${\mathfrak H}_1$ fixed:
\begin{eqnarray}\label{300}
{\cal G}\Pi _1=\Pi _1
\end{eqnarray}

The Frobenius transform commutes with the Fourier transform and also with the projection operators $\pi _r$:
\begin{eqnarray}\label{66}
[F,{\cal G}]=[\pi _r, {\cal G}]=0;\;\;\;\;\;\;r =0,1,2,3
\end{eqnarray}
This can be proved using Eqs.(\ref{65}),(\ref{710}).

\subsection{Frobenius transform in ${\mathfrak H}_d$}
The Frobenius operator in the space ${\mathfrak H}_d$ (where $d$ is a divisor of $\ell$)
is ${\cal G}\Pi _d$.
The analogue of ${\cal G}^\ell ={\bf 1}$ in Eq.(\ref{307}) is here 
\begin{eqnarray}\label{500}
{\cal G}^d\Pi _d=\Pi _d
\end{eqnarray}
The
\begin{eqnarray}
{\rm Gal}(H/{\mathfrak H}_d)=\{{\bf 1}, {\cal G}^d,...,{\cal G}^{\ell -d}\} 
\end{eqnarray}
form a cyclic group of order $\ell/d$ which is a subgroup of ${\rm Gal}(H/{\mathfrak H}_1)$.
These transformations leave all the functions in ${\mathfrak H}_d$ fixed.

We have explained earlier that with a unitary transformation $U$ we go to a different basis
and the subspace ${\mathfrak H}_d$ becomes $U{\mathfrak H}_d$ and the Frobenius transform becomes $U{\cal G}U^{\dagger}$.
In this case
\begin{eqnarray}\label{tt6}
{\rm Gal}(H/U{\mathfrak H}_d)=\{{\bf 1}, U{\cal G}^dU^{\dagger},...,U{\cal G}^{\ell -d}U^{\dagger}\} 
\end{eqnarray}

\subsection{Spectrum of Frobenius transforms}
A direct consequence of Eq.(\ref{307}) is that we can express ${\cal G}$ in terms of its eigenvalues and eigenvectors as:
\begin{eqnarray}\label{37nn}
&&{\cal G}=\varpi _0+\Omega \varpi _1+...+\Omega ^{\ell -1}\varpi _{\ell -1};\;\;\;\;\;\;\Omega= \exp \left (i\frac{2\pi}{\ell}\right )\nonumber\\
&&\varpi _{\lambda }\varpi _{\mu}=\varpi _{\lambda }\delta (\lambda, \mu);\;\;\;\;\;\;\;\sum _{\lambda }\varpi _{\lambda }={\bf 1};\;\;\;\;\;
\lambda,\mu=0,...,\ell-1 
\end{eqnarray}
Here $\varpi _{\lambda }$ are orthogonal projectors to the eigenspaces corresponding to the various eigenvalues of ${\cal G}$.
From Eq.(\ref{37}) it follows that:
\begin{eqnarray}
\varpi _{\lambda }=\frac{1}{\ell}\left [{\bf 1}+(\Omega ^{-\lambda}{\cal G})+(\Omega ^{-\lambda}{\cal G})^2+...+(\Omega ^{-\lambda}{\cal G})^{\ell -1}\right ]
\end{eqnarray}
and using Eq.(\ref{66}) we show that
\begin{eqnarray}
[F,\varpi _{\lambda }]=[\pi _r,\varpi _{\lambda }]=0;\;\;\;\;\;\;r=0,1,2,3;\;\;\;\;\;\;\lambda =0,...,\ell-1 
\end{eqnarray}

Using Eq. (\ref{300}) we prove that 
\begin{eqnarray}
\Pi _1\varpi _0=\varpi _0 \Pi _1=\Pi _1
\end{eqnarray}
It is seen that the space ${\mathfrak H}_1$ is a subspace of the eigenspace of ${\cal G}$ corresponding to the eigenvalue $1$.

More generally, for $d$ which is a divisor of $\ell$, we express ${\cal G}^d$ as
\begin{eqnarray}
{\cal G}^d&=&[\varpi _0+\varpi _{\ell /d}+...+\varpi _{(d-1)\ell/d}]
+\Omega ^d[\varpi _1 +\varpi _{1+\ell /d}+...+\varpi _{1+(d-1)\ell/d}]+...\nonumber\\
&+&\Omega ^{\ell -d}[\varpi _ {(\ell -d)/d}+\varpi _{(2\ell -d)/d}+...+\varpi _{\ell -1}]
\end{eqnarray}
and using Eq. (\ref{500}) we prove that 
\begin{eqnarray}
\Pi _d[\varpi _0+\varpi _{\ell /d}+...+\varpi _{(d-1)\ell/d}]=[\varpi _0+\varpi _{\ell /d}+...+\varpi _{(d-1)\ell/d}]\Pi _d=\Pi _d
\end{eqnarray}
It is seen that the space ${\mathfrak H}_d$ is a subspace of the combined eigenspace of ${\cal G}$ corresponding to the eigenvalues 
$1,\Omega ^{\ell/d},...,\Omega ^{(d-1)\ell/d}$.

In the appendix we present the matrices $\varpi _{\lambda }$ for an example.

\section{Displacements and the Heisenberg-Weyl group}

The displacement matrices are defined as
\begin{eqnarray}\label{412}
&&Z^ {\alpha}(n,m)=\omega [{\rm Tr}(\alpha m)]\;\delta (n,m);\;\;\;\;\;X^\beta (n,m)=\delta (n,m+\beta)\nonumber\\
&&Z^ {\alpha}X^\beta=X^\beta Z^ {\alpha} \omega [{\rm Tr}(\alpha \beta)];\;\;\;\;\;\;\alpha, \beta \in GF(p^\ell)
\end{eqnarray} 
Acting with them on functions $\chi (m)$ in the space $H$ we get
\begin{eqnarray}\label{412a}
&&(Z^ {\alpha}\chi) (n)=\omega [{\rm Tr}(\alpha n)]\;\chi (n);\;\;\;\;\;(X^\beta \chi) (n)=\chi (n-\beta)
\end{eqnarray} 
Both $Z$ and $X$ are $p^\ell \times p^\ell$ matrices with elements which are complex numbers.
We have explained in \cite{5-1} that powers of these matrices to elements of a Galois field, are {\bf defined} through the above equations and they
form the Heisenberg-Weyl group. 

General displacement operators are given by
\begin{eqnarray}\label{90}
&&D(\alpha, \beta)=Z^\alpha X^\beta \omega\left [-\frac{1}{2}{\rm Tr}(\alpha \beta)\right ];\;\;\;\;\;[D(\alpha, \beta)]^{\dagger}=
D(-\alpha, -\beta)\nonumber\\
&&D(\alpha, \beta)D(\gamma ,\delta)=D(\alpha +\gamma ,\beta +\delta)\;\omega \left [\frac{1}{2}{\rm Tr}(\alpha \delta -\beta \gamma)\right ]
\end{eqnarray}
We use the notation $[D(\alpha, \beta)](n,m)$ for the $(n,m)$ element of the matrix $D(\alpha, \beta)$.
They are given by:
\begin{eqnarray}\label{187}
[D(\alpha, \beta)](n,m)=\omega\left [{\rm Tr}(2^{-1/2}\alpha \beta +\alpha m)\right ]\delta (n,m+\beta)
\end{eqnarray}

We can show that 
\begin{eqnarray}\label{414}
&&FD(\alpha, \beta)F^{\dagger}=D(\beta , -\alpha)\\\label{415}
&&{\cal G}^\lambda D(\alpha, \beta)({\cal G}^{\dagger})^{\lambda}=D(\alpha ^{p^ \lambda}, \beta ^{p^ \lambda});\;\;\;\;\;\lambda =0,...,\ell -1
\end{eqnarray} 
Eq.(\ref{415}) has no analogue in harmonic analysis on the field of real numbers or on ${\mathbb Z}_p$.
It is tempting to interpret it as magnification of the phase space, where the `coordinates' are replaced by 
their powers. We stress however that we are in a finite field and magnification is simply reordering.

If $\alpha , \beta$ belongs to the subfield ${\mathbb Z}_p$ then
\begin{eqnarray}\label{416a}
\alpha , \beta\in {\mathbb Z}_p\;\rightarrow\;
{\cal G} D(\alpha, \beta){\cal G}^{\dagger}=D(\alpha, \beta )
\end{eqnarray} 
It is seen that ${\cal G}$ commutes with the matrices $X$ and $Z$.
Normally when two matrices commute, their powers also commute.
${\cal G}$ commutes with `ordinary powers' of $X$ and $Z$ which belong in ${\mathbb Z}_p$;
but it does {\bf not} commute with `extraordinary powers' of $X$ and $Z$
which belong in $GF(p^\ell)-{\mathbb Z}_p$ and which are
are defined through Eq.(\ref{412}). We discuss further this point below in connection with the spectrum of the displacement operators.

A more general result than Eq.(\ref{416a}) is that
if $\alpha , \beta$ belongs to the subfield $GF(p^d)$ then
\begin{eqnarray}\label{416}
\alpha , \beta\in GF(p^d)\;\rightarrow\;
{\cal G}^ {d} D(\alpha, \beta)({\cal G}^{\dagger})^{d}=D(\alpha, \beta )
\end{eqnarray}

We have shown in \cite{5-1} that 
the displacement operators acting on $H$ are expressed in terms of
the displacement operators 
\begin{eqnarray}\label{90}
{\cal D}(\alpha _i,\beta _i)= {\cal Z}^\alpha _i{\cal X}^\beta _i
\omega\left [-\frac{1}{2}\alpha _i\beta _i\right ];\;\;\;\;\;\alpha _i,\beta _i \in {\mathbb Z} _p 
\end{eqnarray}
acting on the various component spaces ${\cal H}$ as:
\begin{eqnarray}\label{p11}
D(\alpha, \beta)&=&{\cal D}({\overline \alpha}_0,\beta _0)\otimes... \otimes{\cal D}({\overline \alpha}_{\ell -1},\beta _{\ell -1})
\end{eqnarray}
The dual components of $\alpha$ and the components of $\beta$, enter in this equation.
We see here another difference between harmonic analysis on $GF(p^\ell)$ and harmonic analysis on
${\mathbb Z}_p\times ...\times {\mathbb Z}_p$. Displacements in the latter will be
${\cal D}(\alpha_0,\beta _0)\otimes... \otimes{\cal D}(\alpha_{\ell -1},\beta _{\ell -1})$.

Special cases of Eq.(\ref{p11}) are
\begin{eqnarray}\label{p11a}
Z^ \alpha=Z^{{\overline \alpha}_0}\otimes... \otimes Z^{{\overline \alpha}_{\ell -1}};\;\;\;\;\;\;\;
X^ \beta= X^{\beta _0} \otimes... \otimes X^{\beta _{\ell -1}}
\end{eqnarray}

\subsection{Properties of displacement operators}

An arbitrary operator $\Theta$ acting on $H$ can be expanded in terms of the displacement operators as
\begin{eqnarray}\label{950}
\Theta=\frac{1}{p^\ell}\sum _{\alpha, \beta} D(\alpha, \beta){\cal W}(-\alpha ,-\beta);\;\;\;\;\;{\cal W}(\alpha ,\beta)={\rm tr}[\Theta D(\alpha, \beta)]
\end{eqnarray}
where tr denotes the usual trace of a matrix.
This is proved using Eq.(\ref{187}). ${\cal W}(\alpha ,\beta)$
is the Weyl (or ambiguity) function.
An important special case is the $SU(p^\ell)$ unitary transformations.
For infinitesimal $SU(p^\ell)$ transformations $\theta$
\begin{eqnarray}
\theta={\bf 1}+\sum _{\alpha, \beta} D(\alpha, \beta) {\mathfrak e}(\alpha ,\beta)
\end{eqnarray}
where ${\mathfrak e}(\alpha ,\beta)$ are infinitesimal coefficients.
In this case the $D(\alpha, \beta)$ (with $(\alpha, \beta)\ne (0,0)$)
play the role of the $p^{2\ell}-1$ generators of the $SU(p^\ell)$\cite{z}.
Finite $SU(p^\ell)$ transformations are given by Eq.(\ref{950}).

Another property is the `generalized resolution of the identity'.
For an arbitrary operator $\Theta$ 
\begin{eqnarray}\label{531}
\frac{1}{p^\ell}\sum _{\alpha, \beta} D(\alpha, \beta)\;\frac {\Theta}{{\rm tr} \Theta}\;[D(\alpha, \beta)]^{\dagger}={\bf 1}
\end{eqnarray}
This is also proved using Eq.(\ref{187}).
There is analogous property in the theory of coherent states for the harmonic oscillator (e.g.,\cite{rev}).
As special case  we use
\begin{eqnarray}
\Theta (m,n)=\psi(m)[\psi (n)]^*;\;\;\;\;\;\;\sum _m|\psi(m)|^2=1
\end{eqnarray}
in Eq.(\ref{531}), where $\psi (m)$ is an arbitrary normalized vector in $H$.
In this case we show that the $p^{2\ell}$ vectors  $D(\alpha, \beta)\psi$ (for all $\alpha, \beta \in GF(p^\ell)$)
form an `overcomplete basis' of vectors in the $p^\ell$ dimensional space $H$.
Indeed Eq.(\ref{531}) shows that we can expand an arbitrary vector $\chi (m)$ as: 
\begin{eqnarray}
\chi (m)=\frac{1}{p^\ell}\sum _{\alpha, \beta} u(\alpha, \beta) [D(\alpha, \beta)\psi](m);\;\;\;\;\;\;
u(\alpha, \beta)=(D(\alpha, \beta)\psi,\chi)
\end{eqnarray}

Other important properties of the displacement operators are the marginal properties\cite{3}:
\begin{eqnarray}\label{789}
\frac{1}{p^{\ell}}\sum _{\alpha \in GF(p^{\ell})}D(\alpha, \beta)&=&{\cal Q}_{-2^{-1}\beta} \nonumber\\
\frac{1}{p^{\ell}}\sum _{\beta \in GF(p^{\ell})}D(\alpha, \beta)&=&{\tilde {\cal Q}}_{2^{-1}\alpha}
\end{eqnarray}
where ${\tilde {\cal Q}}_k$ are projection operators which are the Fourier transforms of ${\cal Q}_k$:
\begin{eqnarray}
{\tilde {\cal Q}}_k=F {\cal Q}_kF^{\dagger}
\end{eqnarray}

\subsection{Displacements and the Heisenberg-Weyl group in ${\mathfrak H}_d$}

In analogy with Eq.(\ref{412}) introduce the   
displacement matrices ${\mathfrak Z}$ and ${\mathfrak X}$ in the space ${\mathfrak H}_d$:
\begin{eqnarray}\label{412b}
&&{\mathfrak Z}^ {\alpha}(n,m)=\omega [{\mathfrak Tr}(\alpha m)]\;\delta (n,m);\;\;\;\;\;{\mathfrak X}^\beta (n,m)=\delta (n,m+\beta)\nonumber\\
&&{\mathfrak Z}^ {\alpha}{\mathfrak X}^\beta={\mathfrak X}^\beta {\mathfrak Z}^ {\alpha} \omega [{\mathfrak Tr}(\alpha \beta)]
;\;\;\;\;\;\;\alpha, \beta \in GF(p^\ell)\nonumber\\
&&{\mathfrak F}{\mathfrak Z}^\alpha {\mathfrak F}^{\dagger}={\mathfrak X}^{-\alpha};\;\;\;\;\;\;
{\mathfrak F}{\mathfrak X}^\alpha {\mathfrak F}^{\dagger}={\mathfrak Z}^{\alpha}
\end{eqnarray} 
We note that the trace of eq.(\ref{776}) is used here.
General dispalcements are defined as
\begin{eqnarray}
&&{\mathfrak D}(\alpha, \beta)={\mathfrak Z}^\alpha {\mathfrak X}^\beta \omega\left [-\frac{1}{2}{\mathfrak Tr}(\alpha \beta)\right ]
;\;\;\;\;\;\;\;\;\alpha, \beta \in GF(p^d)
\end{eqnarray}
${\mathfrak D}(\alpha, \beta)$ is a $p^d \times p^d$ matrix with elements which are complex numbers.
We compare the matrices ${\mathfrak D}(\alpha, \beta)$ with their counterparts $D(\alpha, \beta)$. 
We find that for $\alpha , \beta , n,m\in GF(p^d)$:
\begin{eqnarray}\label{rty}
[\Pi _d D(\alpha, \beta)\Pi_d](n,m)=\left \{[{\mathfrak D}(\alpha, \beta)](n,m)\right \}^{\ell/d}
\end{eqnarray}
This shows that the elements of $D(\alpha, \beta)$ (with $\alpha, \beta \in GF(p^d)$) which have indices $(n,m)$
in $GF(p^d)$ are powers of the corresponding elements of ${\mathfrak D}(\alpha, \beta)$.
Of course, the matrices $D(\alpha, \beta)$ (with $\alpha, \beta \in GF(p^d)$) have other elements also, with indices $(n,m)$ in the set 
$GF(p^\ell)-GF(p^d)$. Furthermore, there are more matrices $D(\alpha, \beta)$ with $\alpha, \beta$ in the set $GF(p^\ell)-GF(p^d)$,
which do not enter in Eq.(\ref{rty}).

\subsection{Spectrum of displacement operators: an example}\label{ht}
$Z$ and $X$ are $p^\ell \times p^\ell$ matrices with $p$ distinct eigenvalues (the powers of $\omega$).
Therefore there is a large degeneracy.
For simplicity we discuss the spectrum of these matrices for the example of the field 
$GF(9)$ ($p=3$ and $\ell =2$). 
The elements of this field are $m_0 +m_1\epsilon$ where $m_0,m_1 \in {\mathbb Z}_3$. They are defined modulo an irreducible polynomial   
which we choose to be $\epsilon ^2+\epsilon +2$.
Below we present matrices using the following order for their indices which are elements of $GF(9)$:
\begin{eqnarray}
\{0, 1, 2, \epsilon, 1+\epsilon, 2+\epsilon, 2\epsilon, 1+2\epsilon, 2+2\epsilon\}
\end{eqnarray}
We consider as an example, the operator ${\cal Q}_2$ defined in Eq.(\ref{900}) and we use Eq.(\ref{950}) to expand it as
\begin{eqnarray}\label{800}
{\cal Q}_2=\frac{1}{9}\sum _{\alpha \in GF(9)}\omega [-2\;{\rm Tr} \alpha]Z^ \alpha
\end{eqnarray}
$Z$ is the $9\times 9$ diagonal matrix
\begin{eqnarray}\label{864}
Z&=&{\rm diag}(1\;\;\;\;\omega ^{2}\;\;\;\;\omega \;\;\;\;\omega ^{2}\;\;\;\;\omega \;\;\;\;1\;\;\;\;\omega\;\;\;\;1\;\;\;\;\omega ^{2})\nonumber\\
\omega &=&\exp \left (i\frac{2\pi}{3}\right )
\end{eqnarray}
Its eigenvalues are $1$, $\omega $ and $\omega ^{2}$ and 
we call $q_0$, $q_1$ and $q_2$ the projection operators to the 3-dimensional eigenspaces corresponding to these eigenvalues.
Then
\begin{eqnarray}
&&Z=q_0+\omega q_1+\omega ^2 q_2\nonumber\\
&&q_0={\cal Q}_0+{\cal Q}_{2+\epsilon}+{\cal Q}_{1+2\epsilon}\nonumber\\
&&q_1={\cal Q}_2+{\cal Q}_{1+\epsilon}+{\cal Q}_{2\epsilon}\nonumber\\
&&q_2={\cal Q}_1+{\cal Q}_{\epsilon}+{\cal Q}_{2+2\epsilon}
\end{eqnarray}
It is clear from this that sums of `ordinary powers' of $Z$ (i.e., powers in the field ${\mathbb Z}_3$)
will give a combination of $q_0$, $q_1$ and $q_2$ and they will never give ${\cal Q}_2$.
In general, the purpose of going to larger fields is to overcome restrictions which we get in smaller fields.
In the present context, powers of $Z$ in $GF(9)$ (defined through Eq.(\ref{412})), are able to give ${\cal Q}_2$, as seen in Eq.(\ref{800}).
In the same spirit, we write the matrix $Z^\epsilon$ as
\begin{eqnarray}
&&Z^ \epsilon ={\mathfrak Q}_0+\omega {\mathfrak Q}_1+\omega ^2 {\mathfrak Q}_2\nonumber\\
&&{\mathfrak Q}_0={\cal Q}_0+{\cal Q}_{\epsilon}+{\cal Q}_{2\epsilon}\nonumber\\
&&{\mathfrak Q}_1={\cal Q}_2+{\cal Q}_{2+\epsilon}+{\cal Q}_{2+2\epsilon}\nonumber\\
&&{\mathfrak Q}_2={\cal Q}_1+{\cal Q}_{1+\epsilon}+{\cal Q}_{1+2\epsilon}
\end{eqnarray}
It is seen that $Z^\epsilon$ is a combination of the projectors ${\mathfrak Q}_0$, $ {\mathfrak Q}_1$, ${\mathfrak Q}_2$
which are different from the projectors $q_0$, $q_1$, $q_2$.
Therefore the degeneracy in the matrices $Z$ and $X$ which might be viewed as an obstacle for expansions like Eq.(\ref{950}),
is not obstacle when we work with powers of these matrices in a larger field.

\section{Symplectic transformations and the $Sp(2,GF(p^\ell))$ group}

General symplectic transformations $S(r,s,t)$ perform by definition, the following unitary transformations:
\begin{eqnarray}\label{sympl}
&&S(r,s,t)Z^ \alpha [S(r,s,t)]^{\dagger}=D(u \alpha, t\alpha)\equiv (Z')^ \alpha
\nonumber\\
&&S(r,s,t)X ^\beta S[(r,s,t)]^{\dagger}=D(s \beta, r \beta)\equiv (X')^\beta \nonumber\\
&&ru-st=1;\;\;\;\;\;\;r,s,t,u \in GF(p^\ell)
\end{eqnarray}
These transformations preserve Eqs (\ref{412}):
\begin{equation}\label{34}
(X')^\beta (Z')^\alpha= (Z')^\alpha (X')^\beta \omega[-{\rm Tr}(\alpha \beta)];\;\;\;\;\;\;\;\;\;\alpha, \beta \in GF(p^\ell)
\end{equation}
The $X'$ and $Z'$ play the same role as the $X$ and $Z$ but in different directions in the $GF(p^\ell)\times GF(p^\ell)$
phase space (which is a finite geometry).

More generally, the displacement operators $D(\alpha ,\beta)$ are trasformed as follows:
\begin{eqnarray}\label{sympl67}
S(r,s,t)D(\alpha ,\beta) [S(r,s,t)]^{\dagger}=D(u\alpha +s\beta , t\alpha +r\beta)\equiv D'(\alpha ,\beta)
\nonumber
\end{eqnarray}
It is seen that displacements by $(u\alpha +s\beta , t\alpha +r\beta)$ in the `old frame' are 
also displacements by $(\alpha ,\beta)$ in the `new frame'.

The transformations (\ref{sympl}) contain three independent variables; and the fourth variable is defined through the constraint.
Since the variables belong to a field, for a given triplet $r,s,t$ (with $r \ne 0$)
there exist $u =r ^{-1}(st +1)$ which satisfies the constraint.
These transformations form the $Sp(2,GF(p^ \ell))$ group.

Following \cite{3} we present the symplectic operators $S(r,s,t)$.
We first give the matrix elements of three important special cases of symplectic operators.
We also explain briefly some of their properties.
The elements of the first one are:
\begin{eqnarray}\label{T1}
&&[S(\xi , 0,0)](n,m)=\delta (\xi ^{-1}n,m);\;\;\;\;\;\;\;S(\xi _1, 0,0)S(\xi _2 , 0,0)=S(\xi _1\xi _2, 0,0)\nonumber\\
&&[S(\xi , 0,0)]^{p^\ell}= [S(\xi , 0,0)]
\end{eqnarray} 
These operators form a subgroup of $Sp(2,GF(p^\ell))$.
 
The second special case of symplectic operators is 
\begin{eqnarray}\label{T2}
&&[S(1,\xi ,0)](n,m)=\omega [{\rm Tr}(2^{-1} \xi m^2)]\delta(n,m);\;\;\;\;\;\;\;S(1,\xi _1, 0)S(1,\xi _2 , 0)=S(1,\xi _1+\xi _2, 0)\nonumber\\
&&[S(1,\xi ,0)]^p= {\bf 1}
\end{eqnarray} 
These operators also form a subgroup of $Sp(2,GF(p^\ell))$.

The third special case of symplectic operators is 
\begin{eqnarray}\label{T3}
&&S(1,0, \xi )=FS(1,\xi ,0)F^{\dagger};\;\;\;\;\;\;[S(1,0,\xi )](n,m)=p^{-\ell}\sum _{\kappa }\omega 
[{\rm Tr}(2^{-1} \xi \kappa ^2 +\kappa n -\kappa m)]\nonumber\\
&&S(1,0, \xi _1 )S(1,0,\xi _2)=S(1,0,\xi _1+\xi _2 );\;\;\;\;\;\;\;
[S(1,0,\xi )]^p= {\bf 1}
\end{eqnarray} 
These operators also form a subgroup of $Sp(2,GF(p^\ell))$.

The general symplectic operator $S(r,s,t)$ is given by
\begin{eqnarray}\label{RRR}
S(r,s,t)&=&S(1,0,\xi_1)S(1,\xi _2,0)S(\xi _3,0,0)\nonumber\\
\xi _1&=&rt (1+st)^{-1} \nonumber\\
\xi _2&=&sr^{-1} (1+st) \nonumber\\ 
\xi _3&=&r(1+st)^{-1}
\end{eqnarray} 
Taking into account Eqs(\ref{T1}),(\ref{T2}),(\ref{T3}) we calculate the matrix elements of  $S(r,s,t)$:
\begin{eqnarray}\label{RRR1}
&&[S(r,s,t)](n,m)=p^{-\ell }\;G(A)\;\omega [{\rm Tr}\;B];\;\;\;\;\;\;\;G(A)=\sum _{k \in GF(p^\ell)}\omega [{\rm Tr}\;(Ak^2)]\nonumber\\
&&A=-2^{-1}(1+st)^{-1}rt;\;\;\;\;\;\;\;B=(2rt)^{-1}[(1+st)n^2-2nmr +m^2r^2]
\end{eqnarray} 
$G(A)$ is the Gauss sum related to $GF(p^\ell)$\cite{g}.

We can show that
\begin{eqnarray}\label{414f}
{\cal G}^\lambda S(r,s,t)({\cal G}^{\dagger})^{\lambda}=S(r ^{p^ \lambda}, s ^{p^ \lambda}, t ^{p^ \lambda})
;\;\;\;\;\;\lambda =0,...,\ell -1
\end{eqnarray} 
If $\alpha , \beta$ belong to the subfield  $GF(p^d)$ then
\begin{eqnarray}\label{416t}
\alpha , \beta\in GF(p^d)\;\rightarrow\;
{\cal G}^ {d} S(r,s,t)({\cal G}^{\dagger})^{d}=S(r,s,t)
\end{eqnarray}

We have seen in Eq.(\ref{ppp1}) that there is a simple relation between Fourier transforms 
in $H$ and Fourier transforms in the component spaces ${\cal H}$.
The same is true about displacements in Eq.(\ref{p11}). 
There is {\bf no} simple relation between symplectic transforms 
in $H$ and symplectic transforms in the component spaces ${\cal H}$.
In order to explain this, we point out that Fourier transforms and displacements involve addition in $GF(p^\ell)$
and the trace of a product.
Addition in $GF(p^\ell)$ is simply addition of $\ell$-dimensional vectors;
and the trace of a product is expressed in a simple way in terms of the dual components in Eq.(\ref{102}).
Symplectic transformations involve 
multiplication in $GF(p^\ell)$ (the products $u\alpha$, $t \alpha$, $s\beta$ and $r\beta$ in Eq.(\ref{sympl})), which
is more complicated than addition and is defined modulo the irreducible polynomial of Eq.(\ref{cha}).
This confirms again the difference between  harmonic analysis on
${\mathbb Z}_p\times ...\times {\mathbb Z}_p$ and 
harmonic analysis on $GF(p^\ell)$.
The two are very different from each other, especially in situations 
where the product of two Galois numbers enters (like the symplectic transformations).

We consider an example which shows clearly that there is no simple relation between symplectic transforms 
in $H$ and symplectic transforms in the component spaces ${\cal H}$.
It is the example with $GF(9)$ considered earlier.
As above polynomials are defined modulo $\epsilon ^2+\epsilon +2$.
In this case we show that
\begin{eqnarray}
S(1,1+\epsilon, \epsilon )X ^\epsilon [S(1,1+\epsilon, \epsilon )]^{\dagger}=D(2\epsilon, 1+2\epsilon)
\end{eqnarray} 
Taking into account that $X ^\epsilon ={\bf 1}\otimes {\cal X}$ and also that in the dual basis $2\epsilon= E_0$ we show that
\begin{eqnarray}
D(2\epsilon, 1+2\epsilon)={\cal D}(1,1)\otimes {\cal D}(0,2)
\end{eqnarray} 
Therefore the $S(1,1+\epsilon, \epsilon )$ can not be the tensor product of two symplectic transformations
${\cal S}_1\otimes{\cal S}_2$ acting on ${\cal H}\otimes {\cal H}$
(the ${\cal S}_1 {\bf 1} {\cal S}_1^{\dagger}$ cannot give the ${\cal D}(1,1)$).

We note that results like Eq.(\ref{789}) have been derived with respect to the 
two particular directions in phase space (which we might call position-momentum or time-frequency).
Acting with $S(r,s,t)$ on both sides of Eq.(\ref{789}) we transform them to other directions of the $GF(p^ \ell)\times GF(p^ \ell)$
phase space which is a finite geometry:
\begin{eqnarray}\label{789e}
\frac{1}{p^{\ell}}\sum _{\alpha \in GF(p^{\ell})}D'(\alpha, \beta)&=&{\cal Q}'_{-2^{-1}\beta}\nonumber\\
\frac{1}{p^{\ell}}\sum _{\beta \in GF(p^{\ell})}D'(\alpha, \beta)&=&{\tilde {\cal Q}}'_{2^{-1}\alpha}
\end{eqnarray}
Here
\begin{eqnarray}\label{789et}
{\cal Q}'_{-2^{-1}\beta}\equiv S(r,s,t){\cal Q}_{-2^{-1}\beta} [S(r,s,t)]^{\dagger}\nonumber\\
{\tilde {\cal Q}}'_{2^{-1}\alpha}\equiv S(r,s,t){\tilde {\cal Q}}_{2^{-1}\alpha}[S(r,s,t)]^{\dagger} 
\end{eqnarray}
are projection operators as in Eq.(\ref{900}) but with respect to the `new frame'.
This property shows that the phase space is isotropic.
This isotropy is of course discrete in the present context.

\section{Discussion}

An important aspect of the theory of Galois fields is the relationship of the `large' field with its subfields.
In the present context, we have studied various aspects of harmonic analysis ( and quantum mechanics) on $GF(p^ \ell)$ and its relationship
to harmonic analysis on the subfield $GF(p^ d)$.
The Frobenius transformations of Eqs. (\ref{307}), (\ref{307a}) play a central role in our study
and lead to the Galois groups of Eqs. (\ref{tt5}), (\ref{tt6}). 

We first studied Fourier transforms and expressed in Eq.(\ref{ppp1})
the Fourier transform in $H$ in terms of Fourier transforms in the component spaces ${\cal H}$.
We have also discussed Fourier transforms in ${\mathfrak H}_d$ in Eq(\ref{410a}) and the spectrum of Fourier transforms in Eq.(\ref{37}).

We next studied displacements and expressed in Eq.(\ref{p11})
displacements in $H$ in terms of displacements in the component spaces ${\cal H}$.
The action of Frobenius transformations on displacements has been discussed in Eqs(\ref{415}),(\ref{416a}),(\ref{416}).
An expansion of an arbitrary operator in terms of displacement operators with the Weyl (or ambiguity) functions as coefficients,
has been given in Eq.(\ref{950}).
A generalized resolution of the identity that involves displacements has been given in Eq.(\ref{531}).
The marginal properties of displacements have been given in Eq.(\ref{789}).
We have also discussed displacements in ${\mathfrak H}_d$ in Eq(\ref{412b}). The spectrum of displacements has been discussed for a particular example 
which however exemplifies the general features.

General symplectic transformations have been given in Eq.(\ref{RRR}),(\ref{RRR1}).
The action of Frobenius transformations on symplectic transformations has been discussed in Eqs.(\ref{414f}),(\ref{416t}).
The phase space is isotropic and acting with symplectic operators on various properties we get analogous properties in a different frame.
We have seen an example of this in Eqs.(\ref{789}),(\ref{789e}). 

The work uses algebraic concepts from field extension in the context of quantum mechanics and harmonic analysis.

\section{Appendix}

We consider the example with $GF(9)$ discussed earlier. 
As above polynomials are defined modulo $\epsilon ^2+\epsilon +2$.
We calculate the matrix ${\cal G}(n,m)$ and its eigenvalues and eigenvectors.
There are six eigenvectors corresponding to the eigenvalue $1$ and we call
$\varpi _0$ the corresponding projection operator. There are
three eigenvectors corresponding to the eigenvalue $-1$ and we call $\varpi _1$ the corresponding projection operator.
They are:
\begin{eqnarray}
\varpi _0=\left (
\begin{array}{ccccccccc}
1&0&0&0&0&0&0&0&0\\
0&1&0&0&0&0&0&0&0\\
0&0&1&0&0&0&0&0&0\\
0&0&0&0.5&0&0&0&0&0.5\\
0&0&0&0&0.5&0&0.5&0&0\\
0&0&0&0&0&0.5&0&0.5&0\\
0&0&0&0&0.5&0&0.5&0&0\\
0&0&0&0&0&0.5&0&0.5&0\\
0&0&0&0.5&0&0&0&0&0.5\\
\end{array}
\right);\;\;\;\;\;\;\;
\varpi _1=\left (
\begin{array}{ccccccccc}
0&0&0&0&0&0&0&0&0\\
0&0&0&0&0&0&0&0&0\\
0&0&0&0&0&0&0&0&0\\
0&0&0&0.5&0&0&0&0&-0.5\\
0&0&0&0&0.5&0&-0.5&0&0\\
0&0&0&0&0&0.5&0&-0.5&0\\
0&0&0&0&-0.5&0&0.5&0&0\\
0&0&0&0&0&-0.5&0&0.5&0\\
0&0&0&-0.5&0&0&0&0&0.5\\
\end{array}
\right)  
\end{eqnarray}
According to Eq(\ref{37})
\begin{eqnarray}
{\cal G}=\varpi _0 - \varpi _1;\;\;\;\;\;\;\varpi _0 + \varpi _1={\bf 1};\;\;\;\;\;\;\;\varpi _0 \varpi _1=0
\end{eqnarray}

\end{document}